\documentclass[aps,prd,10pt,
               preprintnumbers,numbers,sort&compress,nofootinbib,
               showpacs]{revtex4}

\usepackage{amsmath}
\usepackage{amssymb}
\usepackage{hyperref}

\newcommand{\exclude}[1]{}

\newcommand{\beq}{\begin{equation}}
\newcommand{\eeq}{\end{equation}}
\newcommand{\bea}{\begin{eqnarray}}
\newcommand{\eea}{\end{eqnarray}}
\newcommand{\nn}{\nonumber}
\def\la{\langle }
\def\ra{ \rangle }
 \newcommand{\junk}[1]{}

\begin{document}

 \title{  Confinement- Deconfinement Phase Transition in Hot and Dense QCD   at  Large N
} 
  \author{Ariel R. Zhitnitsky}
\affiliation{Department of Physics and Astronomy, University of
  British Columbia, Vancouver,  Canada}

\begin{abstract}
We conjecture that  the confinement- deconfinement phase transition in QCD at large number of colors $N$ and $N_f\ll N$ at $T\neq 0$ and $\mu\neq 0$ is triggered by the  drastic change in $\theta$ behavior. 
The conjecture is motivated by   the holographic model of QCD  where 
confinement -deconfinement phase transition 
 indeed happens precisely at the value of temperature $T=T_c$ where $\theta$ dependence 
experiences a sudden change in behavior\cite{Parnachev:2008fy}. The conjecture is  also supported by 
   quantum field theory arguments  when the instanton   calculations  (which trigger the $\theta$ dependence) are under complete theoretical control for  $T>T_c$, suddenly break down  immediately   below $T<T_c$ with sharp changes in  the $\theta$ dependence.
       Finally, the conjecture is supported by a number of numerical lattice results. We employ this conjecture to study  confinement -deconfinement phase transition of dense QCD at large $\mu$ in large $N$ limit  by analyzing the $\theta$ dependence.  We find  that the confinement- deconfinement phase transition   at $N_f\ll N$ happens at very large quark chemical potential $\mu_c\sim \sqrt{N}\Lambda_{QCD}$. This result    agrees with recent findings by McLerran and Pisarski\cite{McLerran:2007qj}. We also speculate on case when $N_f\sim N$.
 
\end{abstract}
 
\pacs{12.38.Aw, 12.38.Lg}
\maketitle

\section{ Introduction}  
 Understanding the phase diagram at nonzero external parameters $T, \mu$   is one of the most difficult problem in QCD. Obviously, this area  is a prerogative of numerical lattice computations.
 However, some   insights  about  the basic features of the phase diagram may  be inferred    by 
 using some analytical approaches. In particular, some qualitative questions  can be 
formulated and answered by considering a theory with 
  large number of colors $N$ or/and flavors $N_f$, see recent papers on then subject
   \cite{Parnachev:2008fy,McLerran:2007qj,Hidaka:2008yy} and references on previous works therein.
  Generically, to study a phase transition one should find an appropriate order parameter. 
  It is easy  to find an order parameter for gluodynamics when light quarks are not present in the system. 
If  massless quarks are introduced into the system, one can study a chiral phase transition and use  the chiral condensate as an order parameter. For massive, but light quarks this is not an option. However, in the limit of very large $N$ one can consider the  free energy as an order parameter.
In confined phase it is order of one, while in deconfined phase it is order of $\sim N^2$.
Small number of flavors $N_f\ll N$ (massless or massive quarks) does not change the basic picture.

We formulate a different criteria for confinement -deconfinement phase transition, and therefore we use a different order parameter to analyze the phase transition.
The new criteria  is based on observation that the deconfined phase transition is always accompanied by  very sharp changes in $\theta$ behavior which represents our basic conjecture. Therefore, in principle, if our conjecture is correct, one can use any order parameter which nontrivially depends on $\theta$ and study this dependence on two sides of the phase transition line.  Very natural question immediately comes into mind: why and how these two different things
(phase transition vs sharp $\theta$ changes ) could be linked? What is the basic motivation for this proposal? First of all, this criteria is motivated by   the observation that in holographic model of QCD  the 
confinement -deconfinement phase transition 
 happens precisely at the value of temperature $T=T_c$ where $\theta$ dependence 
experiences a sudden change in behavior\cite{Parnachev:2008fy}. Secondly,  the proposal  is supported by  the numerical lattice results \cite{Alles:1996nm} -\cite{ Lucini:2005vg}, see also a review article \cite{Vicari:2008jw},  which unambiguously
suggest that the topological fluctuations are strongly suppressed in deconfined phase, and this suppression becomes more severe with increasing $N$. These general features observed in the lattice simulations 
have very simple explanation within  our proposal on the origin of the 
 confinement -deconfinement phase transition, see next section for details.  Finally, 
 our new criteria is based on a physical picture which  can be shortly summarized as follows.
   
  For sufficiently high temperatures $T>T_c$ the instanton gas is dilute 
with density $\sim e^{-\gamma(T)N}$ which implies  a  strong suppression\footnote{ See 
 \cite{Kharzeev:1998kz} 
and references therein for   earlier discussions  on the subject.} of the topological fluctuations
at large $N$ where $\gamma(T)>0$, see below for details on structure of $\gamma(T)-$ function. The calculations in this region are
under complete field theoretical control and the vacuum energy has a nice
analytic behavior $\sim \cos\theta e^{-\gamma(T)N}$ as function of $\theta$. 
At the critical value of temperature, $T=T_c$ where $\gamma(T)$ changes the sign,
 the instanton expansion 
breaks down and 
one should naturally expect  that at $T=T_c$ there should be  a sharp transition in
$\theta$ behavior as  simple formula $\sim \cos\theta$ can only be valid when
the instanton gas is dilute and semiclassical calculations are justified which is obviously not the case
for $T<T_c$. Therefore, it is naturally to 
 associate sharp changes in $\theta$ behavior  with confinement-deconfinement transition,
just as in the holographic model\cite{Parnachev:2008fy}. There is a very narrow window of temperatures in deconfined phase,   $0<{(T-T_c)}/{T_c} \leq {1}/{N}$ when the instanton expansion is not valid. This vicinity of $T_c$  is extremely interesting, see our comments 
about physics in  this region in conclusion. This region  shrinks to a point at  $N=\infty$.
 
The main goal of this paper is to apply  this criteria to the region with  large chemical potential at large $N$
and $N_f\ll N$ and make a  specific prediction on magnitude $\mu_c(T)$ for  confinement-deconfinement transition 
line at large $\mu$ and sufficiently small $T\ll \mu$. 
The corresponding estimation of $\mu_c(T)$  is  based on well-developed 
instanton calculus in deconfined phase where dilute gas approximation is justified.

 The plan of the paper is as follows.
We start in  Section II  by reviewing  recent work \cite{Parnachev:2008fy} on estimation 
$T_c$   using instanton calculus.  We also present a  picture explaining how and why
two apparently different phenomena (sharp changes in $\theta$ and confinement-deconfinement transition) may in fact be tightly linked. 
 In section III we apply the same technique to argue that 
 the confinement- deconfinement phase transition   happens at very large quark chemical potential $\mu_c\sim \sqrt{N}\Lambda_{QCD}$, where $\mu=\mu_B/N$ is already properly
 scaled quark chemical potential. This result    agrees with recent analysis  by McLerran and Pisarski  \cite{McLerran:2007qj}   which was based on fundamentally different starting point. Finally, in section IV we make few comments for  the case when number of flavors $N_f\simeq N$.

 \section{Confinement- Deconfinement  Phase Transition in hot QCD at large $N$.   }
 
 We start with a short review of ref.\cite{Parnachev:2008fy} where the conjecture 
 (that  the confinement-deconfinement phase transition  happens precisely where $\theta$ behavior 
 sharply changes) was implemented for large $N$ QCD at $T\neq 0$.
 Such a sharp transition is indeed observed  in the holographic model of QCD.  From quantum field theory viewpoint such a transition 
 can be understood   as follows. Instanton calculations are under complete theoretical 
 control in the region $T>T_c$ as the instanton density is parametrically suppressed
 at large $N$ in deconfined region\cite{Parnachev:2008fy},
     \beq
  \label{gamma_N}
 V_{\rm inst}(\theta)\sim e^{-\gamma N} \cos\theta,~~~~ \gamma=\Bigl[\frac{11}{3}
 \ln \left(\frac{\pi T}{\Lambda_{QCD}}\right)-1.86\Bigr].
  \eeq
    It is assumed that a higher order corrections may change the numerical 
   coefficients in $\gamma (T)$, but they do not change the structure of eq. (\ref{gamma_N}).
   The critical temperature is determined by condition $\gamma =0$
where  exponentially small expansion parameter  $ e^{-\gamma N}$ suddenly blows up
and becomes exponentially large.
Numerically,
  it happens at 
   \beq
  \label{T_c_N}
  \gamma=\Bigl[\frac{11}{3}  \ln \left(\frac{\pi T_c}{\Lambda_{QCD}}\right)-1.86\Bigr]=0
 ~~~  \Rightarrow ~~~T_c (N=\infty)\simeq 0.53 \Lambda_{QCD},
  \eeq
  where $ \Lambda_{QCD} $ is defined in the Pauli -Villars scheme. 
Our computations are carried out in the regime where the instanton 
density $\sim \exp(-\gamma N) $ is parametrically suppressed  at  any small but finite $\gamma(T)=\epsilon> 0$ when  $N=\infty$.  From eq. (\ref{gamma_N}) one can obtain the following expression for 
instanton density in vicinity of $T>T_c$, 
   \beq
   \label{T}
   V_{\rm inst}(\theta) \sim \cos\theta \cdot e^{-\alpha N\left(\frac{T-T_c}{T_c}\right)}, ~~~~ 1\gg \left(\frac{T-T_c}{T_c}\right)\gg 1/N.
   \eeq
where $\alpha=   \frac{11}{3}$ and $  T_c (N=\infty)\simeq 0.53 \Lambda_{QCD} $ are estimated at one loop level.
 Such a behavior does  imply that the dilute gas approximation is justified even in close vicinity of $T_c$ as long as $\frac{T-T_c}{T_c}\gg \frac{1}{N}$.     
   Therefore, the $\theta$ dependence, which is sensitive to the 
 topological fluctuations is determined by (\ref{T})  all the way down to the temperatures very 
    close to the phase transition point from above, $T=T_c+ O(1/N)$. 
     The topological susceptibility  
 is order of one for $T<T_c$ in confined phase while it vanishes
 $\sim   e^{-\gamma N}\rightarrow 0$ for $T> T_c$ in deconfined phase.
    Non topological quantum fluctuations on the other hand  could be quite large in this region, but they do not effect the structure of eq. (\ref{T}).
    We do anticipate, of course,   that the  perturbative 
   corrections in the instanton background  may change our numerical estimate for $T_c$ and $\alpha$.
 However, we   do  not expect that a qualitative picture of the phase transition   
    may be affected  as a result of these    corrections. We note that the lattice numerical 
   computations 
    \cite{Alles:1996nm} -\cite{ Vicari:2008jw}    do 
suggest that the topological fluctuations are strongly suppressed in deconfined phase immediately above $T_c$, and this suppression becomes more severe with increasing $N$ starting from 
physically relevant case $N=3$. Holographic QCD also supports this picture\cite{Parnachev:2008fy}.
We do not expect any changes in the picture when small number of flavors
$N_f\ll N$ are introduced into the system\footnote{
We have to make the following remark here in order to avoid any confusions later in the text.   In the presence of the massless chiral fermions the $\theta$ dependence   goes away in QCD in both phases: confined as well as deconfined. It is a simple reflection of    the fact that one can redefine
the fermi fields in the chiral limit such that $\theta$ parameter completely disappears from the partition function.    To avoid the
identical vanishing  of $V_{\rm inst}(\theta) $    one can introduce a non-zero  quark mass $m_q\neq 0$.  It does not effect  any of our estimates    
as long as $N_f\ll N$ as all such changes   lead to a sub leading $1/N$ corrections, see item 3 below.
Our goal here is to  study  the coefficient in front of $\cos\theta$    
  in deconfined phase. By such an analysis  we trigger the point  when this coefficient is suddenly blows up, and the $\theta$ dependence must drastically change. The sharp changes of this coefficient $\sim V_{\rm inst}(\theta)$ we identify with complete reconstruction  of the ground state, drastic 
  changes of the relevant gluon configurations, and finally, with   confinement- deconfinement phase transition.  One should also remark here that the assumption made in \cite{Parnachev:2008fy} on non-vanishing chiral condensate in vicinity $T>T_c$ as a holographic model of QCD suggests, 
   is not crucial for our arguments to hold as it leads    to a sub leading $1/N$ correction, see item 3 below.}.

There are three  basic reasons for a generic structure
(\ref{gamma_N},\ref{T_c_N},\ref{T})  to emerge:\\
{\bf 1.} The presence of the exponentially large ``$T-$ independent"  contribution
( e.g. ~$e^{+1.86 N}$ in eq. (\ref{gamma_N})). This term 
   basically describes  the entropy of the configuration. It is due to a number of contributions such as a 
 number of embedding $SU(2)$ into $SU(N)$ etc;\\
 {\bf 2.} The  presence of the ``$T-$ dependent" contribution to  $V_{\rm inst}(\theta)  $ which comes  from  $\int n(\rho) d\rho$ integration, see below (\ref{instanton}).  It   is proportional to
  \beq 
  \label{rho}
     \left(\frac{\Lambda_{QCD}}{\pi T}\right)^{\frac{11}{3}N}=\exp\Bigl[-\frac{11}{3}N
\cdot \ln \left(\frac{\pi T}{\Lambda_{QCD}}\right)\Bigr].
 \eeq
 {\bf 3.} The fermion related contributions such as a  chiral condensate, diquark condensate  or non-vanishing mass term    enter the instanton density as follows $ \sim\la\bar{\psi}\psi\ra^{N_f}\sim e^{N\cdot \left(\kappa\ln  |\la\bar{\psi}\psi\ra|\right)} $. For $\kappa\equiv\frac{N_f}{N}\rightarrow 0$ this term  obviously leads to a sub leading effects $1/N$ in comparison
  with two  main terms in the exponent (\ref{gamma_N}). Therefore, such terms can be neglected as they do not change any estimates
  at $N=\infty$. It is  in accordance with the general arguments suggesting that the fundamental fermions can not change the dynamics of the relevant gluon configurations  as long as $N_f\ll N$. \\
   
The crucial element in this  analysis   is that both leading contributions (items 1 and 2 above) have exponential $e^N$ dependence,
and therefore at $N\rightarrow \infty$ for $T>T_c$ the instanton gas is dilute with density
$e^{-\gamma N}, ~\gamma>0$ which ensures  a  nice $\cos\theta$ dependence (\ref{T}), while  for $T<T_c$ 
the expansion breaks down, and $\theta$ dependence must sharply change at $T<T_c$.  We have identified such sharp changes with first order phase transition.

Once $T_c$ is fixed    one can compute the entire line of the phase transition $T_c(\mu)$  for    relatively small $\mu \ll T_c$ for large but 
finite $N\gg N_f$.
 The result  in the leading loop order    can be presented  as follows\cite{Parnachev:2008fy},
\beq
\label{mu}
T_c(\mu)=T_c(\mu=0)\Bigl[1- \frac{3N_f\mu^2}{4N \pi^2 T_c^2(\mu=0)} \Bigr], ~~ \mu\ll \pi T_c, ~~~ N_f\ll N.
\eeq
As expected, $\mu$ dependence goes away in large $N$ limit in agreement with general large $N$ arguments\cite{Toublan:2005rq}. This formula is in excellent agreement  
with numerical computations  \cite{de Forcrand:2002ci,de Forcrand:2003hx,mu} which  show very  little changes of the critical temperature $T_c$ with $\mu$ for sufficiently small chemical potential.
In particular, even for the case $N_f= 2, ~N=3$ where the expression (\ref{mu}) is not expected  to give a good numerical
estimate, it still works amazingly  well even for $N=3$. Indeed, 
 the result quoted in  \cite{de Forcrand:2002ci}  can be written as 
 $$T_c(\mu)^{lat}=T_c(\mu=0)^{lat}\Bigl[1- 0.500 (67) \frac{\mu^2}{  \pi^2 T_c^2(\mu=0)^{lat}} \Bigr], ~~~ N_f=2, ~~~N=3.$$
It should  be compared with our theoretical prediction (\ref{mu}) for this case
  $$T_c(\mu)^{th}=T_c(\mu=0)^{th}\Bigl[1- \frac{1}{2} \frac{\mu^2}{  \pi^2 T_c^2(\mu=0)^{th}} \Bigr].$$

 The   eq.(\ref{mu}) suggests very slow change of $T_c$ with $\mu$ at large $N$. Such slow variation 
implies that a sufficiently large  changes  of order  one $\Delta T_c \sim \Lambda_{QCD}$ may occur only when chemical potential changes are very large,  $\Delta \mu\sim \sqrt{N} \Lambda_{QCD}$.  In next section we confirm this expectation by a direct computations of $\mu_c (T=0)$ where  we predict that the confinement -deconfinement phase transition happens at very large $\mu_c (T=0) \simeq \sqrt{N} \Lambda_{QCD}$   if $N_f\ll N$.

One more comment on this proposal. Our  conjecture (that  the confinement- deconfinement phase transition in QCD  is triggered by the  drastic change in $\theta$ at the same point $T=T_c$)   implicitly implies that the configurations  which are responsible for sharp  $\theta$ changes
   must also play a significant  role in confined phase at $T<T_c$. On the other hand, 
at $T>T_c$ the dilute instantons completely 
determine the $\theta $ dependence (\ref{T}) while at $T<T_c$ 
 the   small size    instantons    obviously can  not provide confinement \cite{ILM}.
   How can this be consistent with our conjecture that these two things must be linked?
 We note that quark confinement can not be described in the  dilute gas 
approximation, when the instantons and anti-instantons are well
separated and maintain their individual properties (sizes, positions,
orientations), as it happens at large $T>T_c$.  However,   
in strongly coupled theories the instantons and
anti-instantons lose their individual properties (instantons will
``dissociate '')  their sizes become very large and they overlap.  The relevant
description is that of instanton-quarks\footnote{ Instanton quarks originally appeared in 2d models.
  Namely, using an exact accounting and resummation
of the $n$-instanton solutions in $2d~CP^{N-1}$ models,
 the original problem of a statistical instanton
 ensemble  was   mapped unto a $2d$-Coulomb Gas (CG) 
system of pseudo-particles 
with fractional topological charges $\sim 1/N$ \cite{Fateev}.    
This picture leads to the 
elegant explanation of the confinement phase and other important properties of
the $2d~CP^{N-1}$ models \cite{Fateev}.  
Unfortunately,  similar calculations  in  $4d$ gauge theories
is proven to be much more difficult  to carry out  \cite{Belavin}. },
 the quantum objects with fractional topological charges $\pm 1/N$
which become the dominant quasi-particles. 
 The instanton quarks carry, along with fractional topological charges,  
    the fractional $1/N$  magnetic charges 
      which are capable to  propagate  far away from
instantons- parents  being  strongly correlated with each other. For such  
configurations the confinement
is a possible outcome of the dynamics.
It   makes the instanton quarks to become  the perfect candidates
   to serve as  the dynamical  magnetic  monopoles, the crucial element  
   of  the standard  't Hooft and Mandelstam picture for the confinement \cite{Hooft, Hooft1}.
   This basically represents our proposal   for the answer on the question formulated above. 
   One should emphasize that our arguments (that the instantons dissociate into instanton quarks
   in confined phase) are  not based on any semiclassical analysis 
   performed in strongly  coupled regime. Rather,  they are
    based on   the following observation. The dimensionality  of the moduli space of  the relevant statistical ensemble precisely coincides with the corresponding $k$ instanton 
   measure $4Nk$ with $k$ being integer. This ``coincidence" holds for any gauge group G, not limited to $SU(N)$ case   \cite{JZ}. Similar picture  on  dissociation of the instantons into instanton quarks
   in confined phase has been also  recently advocated in  \cite{Unsal:2008ch}, \cite{Diakonov:2007nv}, 
   see our comments on these papers in \cite{Parnachev:2008fy}.
 
 The main lesson of this section can be formulated as follows:   we presented a number of
 arguments   suggesting that our proposal (which relates two naively unrelated phenomena:
 confinement- deconfinement phase transition and drastic changes in  $\theta$ behavior)  is   consistent with all previously known studies. In particular, it   includes:\\
 $\bullet$ lattice computations of topological susceptibility in the vicinity of the phase transition\cite{Alles:1996nm,Lucini:2003zr,Lucini:2004yh,Del Debbio:2004rw,Lucini:2005vg};\\
 $\bullet$ lattice computations of the critical temperature 
 $T_c(\mu)$ as a function of $\mu$ at small $\mu$ \cite{de Forcrand:2002ci,de Forcrand:2003hx,mu}; \\
 $\bullet$ analysis of  the holographic models of QCD in vicinity of the phase transition
 at nonzero temperature \cite{Parnachev:2008fy,Bergman:2006xn,Gorsky:2007bi}. 
   
 In such circumstances when  the outcome which follows     from the basic conjectured principle   agrees with all known results, 
one should naturally try to extend the  corresponding analysis   to the regions in the parametrical space which are presently  not accessible for study by other means. To be more specific, we
 want to analyze the confinement- deconfinement phase transition at very large chemical potential
 $\mu\geq \Lambda_{QCD}$ when   available technique does not allow to perform the  lattice computations. One should also 
 note that presently available holographic models of QCD also can not address this question.
 With this motivation in mind we want to analyze confinement- deconfinement  phase transition at  large chemical potential and compare the obtained results with corresponding analysis \cite{McLerran:2007qj,Hidaka:2008yy} which is based on fundamentally different  starting point.

 \section{Confinement- Deconfinement  Phase Transition  in dense QCD  at  large N.  }
 
 In this section we estimate the value of $\mu_c$ where the instanton expansion breaks
down and therefore, the $\theta$ dependence should experiences a sharp change. According to our conjecture we should
identify this place with the phase transition point.
Similar arguments have been put forward  previously \cite{Toublan:2005tn} for numerical estimation of  $\mu_c$ for small $N, N_f=2, 3$, see also review talk on this subject\cite{Zhitnitsky:2006sr}.
Our goal here is quite different: we want to understand an analytical dependence of  $\mu_c (N, N_f)$ 
as a function of $N, N_f$ at very large $N$ and finite $N_f\ll N$ in order to compare 
  with results   of refs.
 \cite{McLerran:2007qj,Hidaka:2008yy}  where the authors presented  a very strong argument  suggesting a very  large $\mu_c\sim \sqrt{N}$ where the phase transition could happen.
 
  We follow the same logic as  in\cite{Parnachev:2008fy}, and study the $\theta$ dependence in
  order to make a prediction about the phase transition point $\mu_c$.
  In the regime $\mu > \mu_c $ the $\theta$ dependence 
 is determined by the dilute instanton gas approximation. We expect that the expansion breaks down only  in close vicinity  of $\mu_c$ at large $N$ as it happens in our previous analysis with phase transition at $T=T_c$. According to the conjecture this point will be identified with 
 confinement- deconfinement  phase transition point $\mu_c$.
 In the present case of  analyzing $\mu_c$ rather than $T_c$ discussed previously \cite{Parnachev:2008fy},
 we do not have any support from the  lattice computations, nor from holographic models.
 Still, the basic governing principle remains the same. Therefore we identify    the point where instanton 
 expansion breaks down (and correspondingly a point where  a simple $\cos\theta$ sharply changes to
  something else) with the point $\mu_c$ where the phase transition happens. 
   In our estimates below 
 we assume that the color superconducting phase is realized in deconfined phase
for all $N$, see e.g. recent review \cite{Alford:2007xm}.
 It is known though that for  extremely large $N=\infty$ one could expect that another phase is more energetically favorable\cite{Deryagin:1992rw}.  Still,  for all reasonably large $N$  
 the color superconducting phase prevails \cite{Shuster:1999tn}. In any case, the difference between the two options would lead to  a sub- leading $1/N$ corrections  as explained in item 3 above.

As we shall see below, the instanton density    in deconfined phase has the following generic behavior,  $\sim \cos\theta\exp{[-N\gamma (\mu) ]}$, where $\gamma(\mu)\sim const.+0(1/N)$ in large $N$ limit.
 Such a behavior 
implies that for any small (but finite) positive $\gamma>0$ the instanton density 
is exponentially suppressed   
and our calculations are under complete theoretical control.
In contrast: at arbitrary small and negative 
  $\gamma<0$ the instanton expansion obviously breaks down, 
 theoretical control is lost
  as an exponential growth   $\sim \exp{(|\gamma| N)}$ for the instanton density makes no sense.
The  $\theta $ behavior must  drastically change at this point. Therefore, the value of $\mu_c$ is  determined by the following condition, 
  \beq
 \label{mu_c}
 \gamma (\mu=\mu_c) =0 ~~~~~  \Longrightarrow  ~~~~~\mu_c= c\Lambda_{QCD}.
 \eeq
 Our goal is to compute the coefficient $c$ by approaching the critical point $\mu_c$ from deconfined side of phase boundary.
 Therefore,   we will be interested in the instanton density in 
the dilute gas regime at $\mu> \mu_c$  where analytical instanton calculations are under control.
 
    As we already  mentioned in footnote 2, pg.2  the $\theta$ dependence   goes away in full QCD in both phases: confined as well as deconfined in the presence of the massless chiral fermions.
 However we are interested in the magnitude   of the instanton contribution $\sim V_{\rm inst}(\theta)$     
  in deconfined phase rather than in $\theta$ dependence of full QCD. Precisely this coefficient   triggers  the point where the instanton expansion  suddenly blows up. The sharp changes in $ V_{\rm inst}(\theta)$ we identify with complete reconstruction  of the ground state, drastic   changes of the relevant gluon configurations, and finally, with   confinement- deconfinement phase transition. To avoid identical vanishing of $ V_{\rm inst}(\theta)$ in the presence of massless fermions one can assume 
  a non zero chiral condensate in deconfined phase, as it has been done for hot matter in\cite{Parnachev:2008fy}
with  motivation from  holographic model of QCD.
Also, one can   assume a  non-vanishing masses $m_q\neq 0$ for the fermions, or 
non-vanishing diquark condensate $\la{\psi}\psi\ra\neq 0$ to avoid identical vanishing 
of $ V_{\rm inst}(\theta)$ for dense matter at large $\mu$. None of these assumptions
 effects any numerical estimates given below in the limit $N\rightarrow \infty,~ N_f\ll N$,
 as all these assumptions
   lead to a sub-leading  effects  $\sim 1/N$ which will be ignored in what follows.
    We shall see in section IV that such kind of assumptions   indeed play a crucial role but only   when $N_f\sim N$.
  
  To be definite,  we assume  that the non-vanishing diquark condensate $\la{\psi}\psi\ra\neq 0$ 
 develops for $\mu>\mu_c$. A precise magnitude of the diquark condensate
 is not  essential  for our calculations as it effects 
 only sub-leading  terms $\sim 1/N$ which will be consistently ignored in what follows.
 The instanton-induced effective action for $N_f$ massless fermions can be easily constructed. In particular, 
  for $N_f=2$ flavors, $u, d$ the corresponding expression takes the following form, 
\cite{tHooft,SVZ,Shuryak_mu,Vainshtein:1981wh,Gross:1980br,shuryak_rev},
\begin{eqnarray}
 \label{inst_vertex}
   L_{\rm inst}&=&e^{-i\theta} \int\!d\rho\, n(\rho) 
  \biggl(\frac{4}{3}\pi^2\rho^3\biggr)^{N_f} \biggl\{
  (\bar u_R u_L)(\bar d_R d_L) + \\
  &+& {3\over32} \biggl[ (\bar u_R\lambda^a u_L)(\bar d_R\lambda^a d_L)
  - {3\over4}(\bar u_R\sigma_{\mu\nu}\lambda^a u_L)
    (\bar d_R\sigma_{\mu\nu}\lambda^a d_L) \biggr]
  \biggr\}  + {\rm H.c.}\nonumber
\end{eqnarray}
We wish to study this problem at nonzero  chemical
potential $\mu$ and  nonzero small temperature  $T\ll \mu$ (to be discussed later in the text).  We use the standard formula
for the instanton density at two-loop order \cite{tHooft,SVZ,Shuryak_mu,Vainshtein:1981wh,Gross:1980br,shuryak_rev}
\begin{eqnarray}
\label{instanton}
 n(\rho)= C_N(\beta_I(\rho))^{2N} \rho^{-5}
 \exp[-\beta_{II}(\rho)]  
 \times  \exp[-(N_f \mu^2 + \frac13 (2N+N_f) \pi^2
 T^2)\rho^2], 
\end{eqnarray}
where
\begin{eqnarray}
  C_N = \frac{0.466 e^{-1.679N} 1.34^{N_f}}{(N-1)!(N-2)! },~~
\beta_I(\rho)&=&-b \log(\rho\Lambda_{QCD}), ~~
\beta_{II}(\rho)=\beta_I(\rho)+\frac{b'}{2b} \log\left(\frac{2
  \beta_I(\rho)}{b}\right),  \nn \\ 
b&=& \frac{11}3 N-\frac23 N_f,~~~~~
 b'=\frac{34}3 N^2-\frac{13}3 N_f
N +\frac{N_f}{N}. \nn
\end{eqnarray}
This formula  contains, of course, the standard instanton classical action $\exp(-8\pi^2/ g^2(\rho))
\sim  \exp[-\beta_{I}(\rho)]  $
which however is hidden as it is   expressed in terms of $\Lambda_{QCD}$
rather than in terms of coupling constant $g^2(\rho)$. The   chemical potential $\mu=\mu_B/N$ in this  expression is already properly normalized quark  chemical potential (rather than baryon chemical potential).
By taking the average of eq.(\ref{inst_vertex}) over the
 state  with nonzero vacuum expectation value for the diquark
 condensate $\la{\psi}\psi\ra\neq 0$ as described in \cite{Toublan:2005tn,ssz}, integrating 
 over $\rho$, and taking large $N$ limit 
  using the standard Stirling formula 
    \beq
  \label{stirling}
  \Gamma (N+1)=\sqrt{2\pi N}N^N e^{-N}\left(1+\frac{1}{12N} +O(\frac{1}{N^2})\right)
  \eeq
 one finds  the following expression
 for the instanton induced potential\footnote{ The diquark condensate in large $N$ limit
 has behavior $\la{\psi}\psi\ra\sim \exp(-\frac{1}{g})\sim \exp(-\sqrt{N})$, see e.g. review
 \cite{Alford:2007xm}. 
 It is still  a  sub-leading $1/\sqrt{N}$ effect in comparison with the main terms (\ref{gamma_mu}). Author thanks an anonymous referee for pointing out
 on this, potentially large, correction.},
   \beq
  \label{gamma_mu}
 V_{\rm inst}(\theta)\sim e^{-\gamma N} \cos\theta,~~~~ \gamma=\Bigl[\frac{11}{6}
 \ln \left(\frac{N_f \bar{\mu}^2}{\Lambda_{QCD}^2}\right)-1.1\Bigr], ~~~\mu^2\equiv N\bar{\mu}^2,
  \eeq
  where we introduced reduced chemical potential  $  \bar{\mu}\equiv \mu/\sqrt{N}$ and neglected all powers $N^p$ in front of $ e^{-\gamma N}$.    The crucial  difference in comparison with similar computation
  at nonzero temperature (\ref{gamma_N})  is emerging  of parameter $\bar{\mu}$
  instead of the original quark chemical potential $\mu\equiv \sqrt{N}\bar{\mu}$.
  It implies that the critical chemical potential where $\gamma$ changes the sign
  (and therefore where the phase transition is expected) 
   is parametrically large $\mu_c\sim \sqrt{N}$ because $\bar{\mu}_c\sim 1$, see below
  for numerical estimates. The origin for this phenomenon can be traced from eq. (\ref{instanton})
  where temperature  dependent factor in the instanton density is proportional to $\sim N$ while chemical potential enters this expression with factor $\sim N_f\ll N$. Therefore, a very  large chemical potential $\mu\sim \sqrt{N}\Lambda_{QCD}$ is required in order to achieve the same effect 
  as   temperature $T\sim \Lambda_{QCD}$.     The physics   of this phenomenon can be explained as follows: at $T\sim 1$ a large number of gluons $\sim N^2$ can get excited
  while at $\mu\sim 1$ only a relatively small number of quarks in fundamental representation $\sim N$ 
  can get excited.  Therefore, it requires a very large chemical potential $\mu^2\sim  {N}$ in order  for fundamental quarks play the same role as gluons do at $T\sim 1$.
  As explained above, the critical chemical potential is determined by condition $\gamma =0$
where  exponentially small expansion parameter  $ e^{-\gamma N}$ at $\mu> \mu_c$ suddenly blows up at $\mu<\mu_c$.
Numerically,
  it happens at 
   \beq
  \label{mu_N}
 \gamma=\Bigl[\frac{11}{6}
 \ln \left(\frac{N_f \bar{\mu}^2}{\Lambda_{QCD}^2}\right)-1.1\Bigr]=0
 ~~~  \Rightarrow  
 ~~~~~ \mu_c (N=\infty)\simeq 1.4\cdot \Lambda_{QCD} \sqrt{\frac{N }{N_f}}, ~~~~ N_f\ll N,
  \eeq
  where $ \Lambda_{QCD} $ is defined in the Pauli -Villars scheme. 
 The topological susceptibility  vanishes
 $\sim   e^{-\gamma N}\rightarrow 0$ for $\mu> \mu_c$ while 
 it must be drastically different for $\mu<\mu_c$ as $\theta$ dependence
 must experience some drastic changes in this region as the instanton expansion breaks down, and therefore simple 
 $\cos\theta$ dependence must be replaced by something else.
 It is very likely that the standard Witten's arguments (valid for the confined phase) 
 still hold  in this region $\mu<\mu_c$ in which case the topological susceptibility
  is order of one. 
 
 The $ \Lambda_{QCD} $ in the Pauli -Villars scheme which enters our formula (\ref{mu_N})    is not well-known numerically.
Therefore, for numerical estimates one can trade $ \Lambda_{QCD} $ in favor of $T_c (N=\infty) $ at $\mu=0$ estimated in\cite{Parnachev:2008fy}, see eq. (\ref{T_c_N}). Therefore, our final numerical estimate for $\mu_c(N=\infty)$ can be presented as
follows, 
  \beq
  \label{mu-final}
 \mu_c (N=\infty)\simeq 2.6\cdot \sqrt{\frac{N }{N_f}}\cdot T_c (N=\infty, \mu=0) , ~~~~ N_f\ll N.
  \eeq
  If one uses the numerical value for $T_c(N=3)\simeq 260$~MeV  
  \cite{Lucini:2003zr,Lucini:2005vg}, one arrives to $\mu_c (N=\infty)\simeq 690  \sqrt{N/N_f}$~MeV  which is our final numerical estimate for the critical chemical potential where deconfined phase transition is predicted for very large $N$. 
Few remarks are in order:\\
{\bf a.}
The most important result of the present studies is the observation that the confinement- deconfinement 
phase transition according to (\ref{mu_N}) happens at very large $\mu_c\sim \sqrt{N}$ if $N_f\ll N$.
This is consistent  with the results of \cite{McLerran:2007qj} where parametrically large scale for $\mu_c\sim \sqrt{N}$ had been predicted.
However, the technique of ref. \cite{McLerran:2007qj} does not  allow to 
answer the question whether the transition would be  the first order   or  it would be  a  crossover.  
Within our framework at $N\gg 1$ and $ N_f\ll N$ the entire   phase transition line  
(which starts at $T=T_c \sim \Lambda_{QCD}$ at $\mu=0$ 
and ends at $\mu=\mu_c\sim \sqrt{N} \Lambda_{QCD}$ at $T=0$)
is predicted to be the first  order phase transition at large $N$ and $N_f\ll N$. This is because the nature  for the phase transition along the entire line  is one and the same: it is drastic changes of $\theta$ dependence when the phase transition line is crossed.\\
{\bf b.} Our computations are carried out in the regime where the instanton 
density $\sim \exp(-\gamma N) $ is parametrically suppressed at $N=\infty$.
From eq. (\ref{gamma_mu}) one can obtain the following expression for 
instanton density in vicinity of $\mu>\mu_c$, 
   \beq
   \label{mu1}
  V_{\rm inst}(\theta) \sim \cos\theta \cdot e^{-\alpha N \left(\frac{\mu-\mu_c}{\mu_c}\right)}, ~~~~ \frac{1}{N}\ll\left(\frac{\mu-\mu_c}{\mu_c}\right)\ll 1,
   \eeq
where $\alpha$ is $11/3$ at one loop level, but the perturbative corrections could be large 
and they may considerably change this numerical coefficient.  
 Such a behavior (\ref{mu1}) does  imply that the dilute gas approximation is justified even in close vicinity of $\mu_c$ as long as $\frac{\mu-\mu_c}{\mu_c}\gg \frac{1}{N}$.    In this case the diluteness parameter  remains small. We can not rule out, of course, the possibility that the  perturbative 
   corrections may change our numerical estimate for $\mu_c$. 
However, we   expect that a qualitative picture of the phase transition advocated in this 
paper remains unaffected  as a result of these   perturbative  corrections   in dilute gas regime.\\
{\bf c.} In our estimate for $\mu_c$   we neglected $(\log \rho\Lambda_{QCD})^k$ 
in evaluating of the $\int d\rho  $ integral.   The corresponding 
correction changes  our estimate (\ref{mu_N}) very slightly, and it will be ignored in what follows. 
Numerical  smallness of correction is   due to the strong  cancellation
between the second loop contribution in the exponent (term proportional to $b'/ b$) and the first loop 
contribution in the pre-exponent in eq. (\ref{instanton}).    \\
   {\bf d.}  Once $\mu_c$ is fixed   one can compute the entire segment of the  phase transition line $\mu_c(T)$  for    relatively small $T$. 
   Indeed, in the dilute gas  regime at $\mu>\mu_c$  the $T$ dependence of the instanton density is determined
by a simple insertion   $\sim \exp[- 2/3 N \pi^2T^2\rho^2]$ in the expression
for the density (\ref{instanton}). In the leading loop order  $\mu_c(T)$ varies as follows,
\beq
\label{T1}
\mu_c(T)=\mu_c(T=0)\Bigl[1- \frac{N \pi^2T^2}{3N_f \mu_c^2(T=0)} \Bigr], ~~~~~~~~~~~
 \sqrt{N}T\ll \mu_c. 
\eeq
One should remark that a variation of the critical chemical potential $\Delta \mu_c(T)$ is very large  $\sim \sqrt{N}$ when the temperature variation   $\Delta T\sim 1$ is  order of one in units of $\Lambda_{QCD}$. This is in huge contrast with a similar expression (\ref{mu}) which shows very little change  $\sim 1/N$ of the critical temperature $\Delta T_c(\mu)\sim 1/N$  with   variation of chemical potential of order one, $\Delta \mu\sim 1 $. 
 The nature    of this difference between $\mu_c$ and $T_c$  was already mentioned before and can be explained by  the fact  that at  $T\sim 1$ a large number of gluons $\sim N^2$ can get excited
  while at $\mu\sim 1$ only a relatively small number of quarks in fundamental representation $\sim N$ 
  can get excited.  Therefore, it requires a very large chemical potential $\mu^2\sim  {N}$ when quarks can   play the same role  as gluons do at  $T\sim 1$ as long as $N_f \ll N$.
  
  \section{Deconfinement   Transition  in hot and dense QCD  at $N_f\sim N$. Speculations.}
  
Our estimations (\ref{mu_N},\ref{mu-final},\ref{T1}) have been derived under assumption that $N\rightarrow \infty$ while $N_f$ is kept  fixed
such that $\kappa\equiv N_f/N\rightarrow 0$.   In particular, for the case of hot matter with $T\neq 0$
  studied  in  \cite{Parnachev:2008fy}
 the fermi fields and the chiral condensate were introduced exclusively with a single  purpose
to elucidate the physical interpretation of the phase transition for 
   pure gluodynamics rather than for full QCD.
The physical results in that case did not depend on $N_f$ nor they depend  on  a magnitude of the chiral condensate in deconfined phase or 
a value of the  quark's mass if it would be  nonzero. The same remark  also applies  for analysis of dense
matter with $\mu\neq 0$ discussed  in the previous section as long as $\kappa\equiv N_f/N\rightarrow 0$.

In this section we want to speculate what happens when $\kappa\equiv N_f/N\sim 1$ by considering
hot matter $T\neq 0, ~\mu\simeq 0$.
\exclude{
   We shall follow the same logic as  before  in order to  study the $\theta$ dependence in
 deconfined phase.
 The basic governing principle remains the same, and  therefore we identify    the place where instanton 
 expansion breaks down (and correspondingly a point where  a simple $\cos\theta$ sharply changes to
  something else) with the point   where the phase transition happens.}
   In contrast with previous analysis
  we anticipate  a very strong dependence  from  all (previously unessential) parameters such as quark's mass $m_q$, number of flavors $N_f=\kappa N$, magnitude of the chiral condensate $\la\bar{\psi}\psi\ra$, etc.  We start by considering
  variation  of transition properties on   
  quark's masses.
For simplicity,  consider the  limit when   all  quarks have the same  and sufficiently large mass $m_q \gg \Lambda_{QCD}$.   It is obvious that the first order phase transition anticipated in pure gluodynamics will not be effected by presence of $N_f$ sufficiently heavy fermions as they essentially decoupled from the system  in the limit $m_q \gg \Lambda_{QCD}$.
  When $m_q$ is getting smaller but still sufficiently large $m_q\geq \Lambda_{QCD}$
  one can easily demonstrate  that the structure of $\gamma(T)$ remains the same, but the corresponding critical temperature will be slowly decreasing with $m_q$ as follows
  \beq
  \label{m}
 T_c(\kappa\neq 0, m_q )=T_c(\kappa=0)\left(1-\kappa\frac{3}{11}\cdot\frac{2}{75}
  (\frac{ \pi T_c}{m_q})^2 -\kappa\frac{3}{11}\cdot \frac{34}{735}(\frac{\pi T_c}{m_q})^4+ ...\right),~~m_q\gg \Lambda_{QCD}, ~~\kappa=\frac{N_f}{N}\sim 1
  \eeq
 where the few first coefficients   of the expansion $1/(m_q\rho)^k$ in   the instanton background
  have  been explicitly calculated long ago ~\cite{Novikov:1983gd},
   see also recent paper \cite{Kwon:2000kf}.
  For our estimates (\ref{m}) we replaced $\rho\rightarrow (\pi T_c)^{-1}$ as a typical value of $\rho$
  where the integral $\int d\rho$ converges. The expansion (\ref{m}) can be trusted   starting from  $(m_q\rho)\sim {m_q}/{(\pi T_c)} >1$. The fermion contribution is a sub leading  effect $\sim1/N$; it becomes of order one when $\kappa=\frac{N_f}{N}\sim 1$, as expected.

 Now, we want to demonstrate a strong dependence of the transition as a function of   the chiral condensate $\la\bar{\psi}\psi\ra$ magnitude.
  If the chiral condensate
  does not vanish identically in close vicinity  of the phase transition, $T>T_c$ as holographic model of QCD suggests  \cite{Parnachev:2008fy},  one can repeat the corresponding calculations  
  with the following result: the structure of $\gamma(T)$ function as defined in (\ref{gamma_N}) remains the same while its coefficients    would  now depend  on dimensionless  parameters $\kappa$ and the value of the  chiral condensate\footnote{Indeed,
  in our previous analysis the chiral condensate enters the instanton density as follows $ \sim\la\bar{\psi}\psi\ra^{N_f}\sim e^{N\cdot \left(\kappa\ln  |\la\bar{\psi}\psi\ra|\right)} $. For $\kappa=\frac{N_f}{N}\rightarrow 0$ this term  obviously leads to a sub leading effects $1/N$ in comparison
  with the main terms in the exponent (\ref{gamma_N}). For $\kappa\sim 1$ this terms becomes the same order of magnitude as other contributions in (\ref{gamma_N}).}.
  The numerical values of the critical temperature $T_c$ and coefficient $\alpha$
  would change, however  the sharp changes of $\theta$ dependence which is a consequence of
  a generic structure of $\gamma(T)$ remain the same. Therefore, we expect the first order phase transition to hold in this case in complete analogy  to   previously considered case $\kappa\rightarrow 0$.  
  
 However,   we think it is very unlikely   for the chiral condensate to remain finite 
    at $T>T_c$ when $\kappa\equiv N_f/N\sim 1$. It is much more likely that 
 the chiral condensate  vanishes    at $T>T_c$ when $\kappa\sim 1$. In this case  our analysis based on   the dilute  instanton approximation (\ref{instanton}) will be obscured due to the long range interactions between  instantons and  anti-instantons 
 induced by massless quarks, \cite{Shuryak:1987an,Velkovsky:1997fe}. 
 This induced interaction becomes crucial 
 even when instantons are still  far away from each other, and the  instanton gas is still dilute.
 The corresponding estimations for $T_c$ and studying the properties of the transition  in the case $\kappa\sim 1$  become very model dependent analysis, and we shall not elaborate on this issue in the present paper.  We anticipate that  the   transition properties will be very sensitive   to the quark's masses as the instanton interactions drastically depend on  the quark's features in this case. Such a sensitivity is consistent with  the  lattice results  which suggest that for vanishing quark masses there will be first order phase transition while for physical masses it becomes a 
 smooth crossover, 
 see  e.g. recent reviews \cite{Kogut:2004su,Stephanov:2004wx}.
   We should emphasize here that our basic principle which relates sharp changes in $\theta$ dependence and   transition properties still holds
   for $\kappa\sim 1$ case. This principle is a simple reflection of the fact that the point where the confinement sets in corresponds to the
   regime where the instanton density suddenly blows up and the instantons dissociate into the instanton quarks as mentioned in chapter II
   and discussed in a more details in \cite{Parnachev:2008fy} and references therein. The large number of fermions when $\kappa\sim 1$
   obscures a simple analysis when the critical point can be estimated by  approaching  from deconfined phase where the instanton density is parametrically suppressed (\ref{T}) and the system remains  under theoretical control up to a close vicinity of $T_c$. In the case of $\kappa\sim 1$
 the corresponding analysis becomes much more involved due to the reasons mentioned above.
  
   Therefore, the main  lesson from 
estimates presented above is as follows. In the case when $N_f\sim N$ we observe a great sensitivity
of the transition properties on specific details of the system such as quark' s masses, magnitude of the chiral condensate, value of $\kappa$. It is very difficult to make any solid  predictions in this situation as they would largely depend
on underlying  assumptions. Such a  sensitivity of the transition to   quark's properties at $\kappa\sim 1$ is in a huge contrast with our previous estimates when 
there is unambiguous  prediction for the first order phase transition  at $T\sim 1$ 
and small chemical potential (\ref{T_c_N},\ref{T},\ref{mu}) and very large $\mu\sim \sqrt{N}$ and small temperature (\ref{mu-final},\ref{mu1},\ref{T1}) 
 when  $\kappa\ll 1$ and when all specific quark's details   are irrelevant as their contribution is suppressed at least by factor $1/\sqrt{N}$.

\section{summary. Future directions } 

We 
explore the consequences of the assumption that in the large $N$ QCD 
at $N_f\ll N$
confinement-deconfinement phase transition
takes place exactly at the point  where the dilute instanton
calculation breaks down, and therefore where  $\theta$ dependence must drastically change.  
This conjecture for $T\neq 0$ 
is supported by lattice computations and holographic arguments. At large chemical potential
  we do not have such independent support. However, the basic governing principle remains the same,
  and therefore, our results (\ref{mu-final},\ref{mu1},\ref{T1})  can be considered as a prediction. 
  The most important consequence of our conjecture  is observation that the critical $\mu_c$ is very large, $\mu_c\sim \sqrt{N}$ which is 
  consistent with fundamentally different arguments presented in ref. \cite{McLerran:2007qj}. Another important   observation is the fact
   that the first order phase transition at $\kappa\ll 1$ holds all the way down from $T_c\sim 1, \mu=0$ to
  $\mu_c\sim \sqrt{N}, T=0$ as a consequence  of the same nature of the transition. In different words,
  the  $\theta$ behavior experience sharp changes whenever the phase transition line is crossed. This feature is very robust consequence of our conjecture, not sensitive to details of quark's properties such as masses, chiral condensation etc, as they may influence  the sub-leading   terms only. 
  Situation becomes drastically different when $N_f\sim N$ in which  case everything becomes very sensitive to details: quark masses, chiral condensate, precise value of $\kappa$, etc.    
  
A  general  comment on this proposal can be formulated as follows.
 Our conjecture which relates two apparently unrelated phenomena (phase transition vs sharp changes in $\theta$ behavior)   implicitly implies that topological configurations which are linked to
 $\theta$ must play a crucial role in the dynamics of the phase transition. For $T>T_c$ such configurations are well-known: they are dilute instantons with density $\sim e^{-\gamma N}\cos\theta$.
 We presented arguments in\cite{Parnachev:2008fy} 
(see also earlier references therein) suggesting that at $T<T_c$ the instantons do not disappear from the system, but rather dissociate into  fractionally charged   constituents, the so-called instanton quarks.   In this sense the phase transition can be understood as a phase transition between molecular phase (deconfined)
  and plasma phase (confined) of these fractionally charged   constituents. The same arguments still hold
  for the entire phase transition line in $(T, \mu)$ plane.
  A similar  conclusion on sharp changes in $\theta$ behavior at $T=T_c$   was also observed  in ref.\cite{Gorsky:2007bi} where
   the authors studied the D2 branes in confined and deconfined phases at $T\neq 0$.
 The topological objects (sensitive to $\theta$)
were identified  in ref.\cite{Gorsky:2007bi} as magnetic strings. 

Our final remark here is as follows. If the picture advocated in the present work about the nature of the transition turns out to be correct, it would strongly suggest  that fractionally charged   constituents (which carry the magnetic charges as discussed in\cite{Parnachev:2008fy} and references therein) may play a very important
  role  in dynamics in deconfined phase in close vicinity of the  transition $0< (T-T_c)\leq 1/N$. 
  For large $N$ 
  this region shrinks to a point, however for finite $N$ it could be an extended region in temperatures.
  In this region the instantons are not formed yet, and our semiclassical analysis is not justified yet
  as eq. (\ref{T}) suggests. However, the constituents in this region are already  not in condensed form. Therefore they may   become an important  magnetic degrees of freedom which may contribute
  to the equation of state, similar to analysis on
  wrapped monopoles in ref.\cite{Chernodub:2006gu,Chernodub:2008vn}.
  The role of these fractional magnetic constituents could be even more profound if  $N_f\sim N$ 
 where smooth crossover likely to take place \cite{Kogut:2004su,Stephanov:2004wx}. In this    case the region of interests is    
 order of one  $(T-T_c)\sim 1$ in $\Lambda_{QCD} $ units in contrast with  a narrow region $0< (T-T_c)\leq 1/N$ if the first order phase transition takes place.    The region above $T_c$ is also very interesting from phenomenological viewpoint  as reviewed in \cite{Shuryak:2008eq}.

\acknowledgments   
I  thank   the FTPI, Minnesota 
for organizing the workshop
 "Continuous Advances in QCD  
May 15-18, 2008". I also thank 
  Larry McLerran for his presentation  at this workshop
 which motivated/initiated  this study\cite{McLerran:2008ux}.
This work 
was supported, in part, by the Natural Sciences and Engineering
Research Council of Canada.

\end{document}